\documentclass[10pt,letterpaper]{article}

\usepackage{cogsci}
\usepackage{graphicx}

\cogscifinalcopy 

\usepackage[
  style=apa,
  natbib=true,
  annotation=false,
]{biblatex}
\usepackage{float} 

\title{From semantic memory to collective creativity: A generative cognitive foundation for social creativity models}

\author[1]{\mbox{Mirza Nayeem Ahmed (ahmed.mir@northeastern.edu)}}
\author[2]{\mbox{Raiyan Abdul Baten (rbaten@usf.edu)}}
\affil[1]{Khoury College of Computer Sciences, Northeastern University}
\affil[2]{Bellini College of Artificial Intelligence, Cybersecurity and Computing, University of South Florida}

\begin{document}

\maketitle

\begin{abstract}
Simulation-based theory development has yielded powerful insights into collective performance by linking social structure to emergent outcomes, yet it has struggled to extend to collective creativity. Creativity is hard to capture purely at the social level, as novel ideas are generated through cognitive mechanisms. To address this gap, we introduce a multi-level socio-cognitive agent-based framework in which agents share a common semantic vocabulary and substrate but differ in semantic network topology. A single generative parameter tunes semantic modularity, yielding emergent individual differences in ideational breadth. When agents exchange ideation traces, two canonical social-creativity phenomena arise without being imposed: lower pre-interaction ideation overlap predicts larger stimulation gains, and shared inspiration sources induce network-level redundancy. The framework enables mechanistic theory-building about cognition and social structure in collective creativity.

\end{abstract}

\section{Introduction}
\label{sec:intro}
Simulation-based theory development has rich precedence in the study of collective performance in complex social systems. Agent-based computational models have clarified how collective outcomes depend on interaction topology, information flow, and exploration-exploitation trade-offs in settings such as collective wisdom and collective problem solving---from classic opinion aggregation models to rugged landscape search~\citep{baten2025wisdom,lazer2007network}. Yet one major class of human performance remains comparatively under-modeled in this tradition: \emph{collective creativity}~\citep{acar2023collective}. Creative outcomes are hard to capture purely at the social level, as idea generation relies on cognitive processes unfolding within individuals. As a result, creativity resists models that operate solely on social interaction rules or assign agents ``creativity scores'' that sever collective dynamics from the cognitive mechanisms that generate ideas.

This gap reflects a non-trivial modeling challenge. We argue that to be useful for mechanistic theory-building, a computational model of collective creativity should bridge cognitive and social levels: \textit{First}, a population of agents must share a common semantic grounding so that an ideation `prompt' refers to the same conceptual object across agents. \textit{Second}, agents must differ in creative ability in a way that \emph{emerges} from cognitive mechanisms rather than being imposed as an exogenous trait. \textit{Third}, social creativity phenomena---such as cognitive stimulation from diverse peers and redundancy from shared sources---should be explainable by the model as a consequence of interaction. Despite extensive empirical work on peer and network effects on creativity and rich cognitive models of individual idea generation, there remains no computational framework that satisfies these constraints jointly.

To address this challenge, we introduce a multi-level, socio-cognitive agent-based model in which agents are endowed with explicit mechanisms for idea generation and embedded in a shared social environment. Strikingly, we find that a \textit{single structural parameter} of semantic memory can serve as a generative ``creativity knob,'' inducing individual creative ability differences across agents. Once these agents interact socially through minimal ideation-trace exchange, canonical social-creativity effects emerge without being imposed.

By moving beyond purely social abstractions and grounding collective creativity in generative cognitive structure, this work provides a minimal yet expressive foundation for modeling creative social systems. We show how creative outcomes can be constructed, varied, and propagated from first principles in a social setting---enabling mechanistic theory-building and controlled counterfactual analysis of collective creativity.

\section{Background}
\label{sec:background}
\subsection{Computational models of collective performance}
Agent-based computational models have been particularly effective when the modeled quantities are clearly specified and the mechanisms of social interaction are explicit. Paradigms such as DeGroot-style opinion dynamics formalize how individuals iteratively update beliefs through social influence~\citep{degroot1974reaching}, while NK landscapes model collective problem solving over rugged payoff spaces~\citep{gomez2019clustering,kauffman1989nk}. These models establish a methodological template: in well-specified settings, simulation can allow inductively examining macro-level impacts of manipulating hypothetical, counterfactual system features~\citep{homer2013complex,lazer2009computational}.

Collective creativity, however, does not trivially fit this template. Early computational attempts at collective creativity models assigned creativity scores to agents with hard-coded redundancy penalties~\citep{baten2020creativity}, or assigned performance outcomes randomly to capture chance effects~\citep{yin2019quantifying}. However, such assignments do not have explicit connection to the underlying cognitive processes, limiting the models' usefulness in theory-building and hypothesis generation about the interactions between socio-cognitive processes.

\subsection{Creative cognition as semantic search}
At the same time, cognitive science offers rich mechanistic accounts of individual creativity~\citep{kenett2023creatively,kumar2021semantic,kenett2018flexibility}. A prominent class of models represents semantic memory as a network of concepts and treats idea generation as a stochastic search process over this structure, such as spreading activation or random walks~\citep{baronchelli2013networks,kenett2016examining,lundin2023neural,collins1975spreading,avery2018comparing,abbott2015random}. The classic associative theory frames novelty as the ability to move beyond obvious, local associations to form non-obvious connections among more remote concepts~\citep{brown2002making,mednick1962associative}. In this view, ideation often begins with easily accessible neighbors of a prompt concept, and novelty emerges when retrieval traverses farther regions of semantic memory and recombines distant elements~\citep{runco2014creativity}.

This perspective has been strengthened by empirical work linking creative ability to properties of semantic organization~\citep{chen2025dynamic,herault2024creative,benedek2017semantic,campidelli2026creativity}. For instance, highly creative individuals enjoy semantic network structures that facilitate access to distant concept regions during retrieval~\citep{kenett2014investigating,siew2019cognitive}. These accounts, however, are predominantly individual-level: they do not address how cognitively grounded agents interact in a shared semantic space, exchange inspiration, or collectively generate novelty. 

\subsection{Peer and social network effects on creativity}
Empirical work across social and network science bodies emphasize that creativity is shaped by exposure to others’ ideas, collaboration structure, and the diversity of social inputs~\citep{baten2021cues,uzzi2005collaboration,vasudeva2013embeddedness,van2012defying,sosa2011creative}. For instance, relationship strength, position and external ties are known to influence creative performance~\citep{perry2006social,porter2020meta}. Non-redundant contacts and cross-boundary interaction (``structural holes'') can provide informational and ideation advantages~\citep{burt2004structural,todo2016strength}. At the same time, shared sources and repeated exposure can lead to redundancy in what people consider and produce~\citep{baten2022novel,kelty2025innovation,zhou2009social}.

Taken together, prior work provides strong socio-cognitive components but lacks an integrative modeling framework. Cognitive models specify how ideas are generated, while social science documents how peer structure shapes creative outputs. What is missing is a generative framework that connects cognitive mechanisms of idea generation to population-level collective ideation dynamics---a gap we address in this paper.

\section{Constructing Creative Agents with Shared Semantic Grounding}
\label{sec:construct}
We first build a population of cognitively grounded agents who can (i) ideate on the same prompts in a common conceptual space, and (ii) differ systematically in creative ability \emph{without} having to be assigned creativity as an exogenous trait. Achieving (i) requires two distinct forms of semantic grounding: a \emph{shared vocabulary} so that a prompt denotes the same concept across agents, and a \emph{shared substrate environment} so that agents are comparable variants of the same semantic universe rather than of unrelated/arbitrary semantic universes. Achieving (ii) requires a structural dimension (a `knob') of semantic memory that is theoretically linked to creative cognition and can be tuned in a controlled, interpretable way.

We represent each agent’s semantic memory as an unweighted, undirected graph $G_i=(V,E_i)$, where nodes $V$ denote concepts and edges encode associative relatedness. Shared vocabulary is enforced by holding the node set $V$ fixed across agents. Shared substrate is enforced by generating agents as structured variants around a common backbone or substrate graph $G_0$ defined on the same $V$. The agents thus differ in their edge sets $E_i$, providing a means to inject individual differences in their semantic structures and subsequent creative outcomes. Our choice of $G_0$ and the approach to introduce idiosyncratic organization in $E_i$ are reasoned next.

\subsection{Design requirement: A structural knob tied to creative cognition}
\label{sec:knob}
A key modeling choice in our work is that we do not assign agents creativity scores. Instead, we vary a \emph{structural property} of semantic memory that prior work associates with creative cognition: semantic modularity. Indeed, recent empirical evidence suggests that more creative individuals tend to exhibit semantic networks with weaker modular organization~\citep{kenett2014investigating,siew2019cognitive}. Intuitively, highly modular semantic networks decompose into densely connected concept-communities with relatively sparse bridges across communities. Such compartmentalization can trap concept retrieval within local neighborhoods during idea generation (diminishing creativity), whereas a weaker modular structure can provide more cross-community connections to support access to remote concept regions (elevating creativity). This makes modularity an attractive candidate for inducing creative heterogeneity without changing how agents search.

Our design requirement is therefore concrete: we need a generative procedure that (a)~yields a family of graphs that are comparable variants of a shared semantic universe (i.e., shared vocabulary $V$ and shared substrate $G_0$), and (b)~provides an interpretable, controllable parameter that systematically shifts the graphs' modularity, $Q(G_i)$ without otherwise modifying the agents' cognitive processes.

\subsection{A shared substrate with a single generative knob}
\label{sec:substrate}
We meet these requirements using a Watts--Strogatz (WS) small-world construction \citep{watts1998collective} as a \emph{generative scaffold}. Prior empirical evidence suggests that semantic networks show small-world properties~\citep{kenett2014investigating,schilling2005small,marupaka2012connectivity}. In particular, the WS family provides a minimal and interpretable continuum between locally clustered neighborhoods and the injection of long-range shortcuts---mimicking the structural intuition that modularity is intended to capture. 

Concretely, we first generate a shared backbone $G_0$ over $n=|V|$ nodes arranged in a ring, where each node is connected to its $k$ nearest neighbors. We then instantiate each agent's semantic network $G_i$ by rewiring each edge of $G_0$ with probability $p_i\in [0,1]$, leading to idiosyncratic associative edges $E_i$. Low $p_i$ preserves local neighborhood structure, yielding strongly compartmentalized graphs; higher $p_i$ injects long-range shortcuts that increasingly blur community boundaries. Importantly, $p_i$ changes what concepts an agent can reach by altering the topology of associations, not by changing the agent's search strategy, effort, or hyperparameters.

\begin{figure}[t]
  \centering
  \includegraphics[width=1\linewidth]{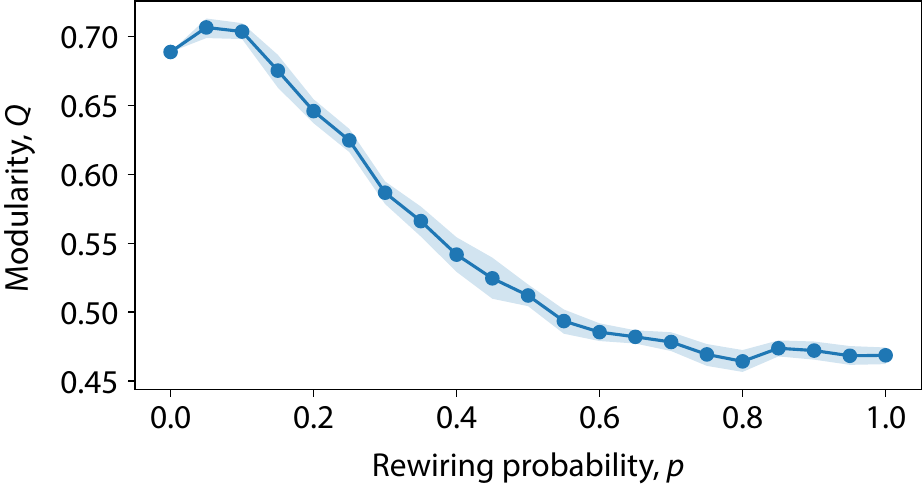}
  \caption{\textbf{Modularity $Q$ decreases as the Watts--Strogatz rewiring probability $p$ increases.} We generate a shared $G_0$ with $|V|=100$ nodes and average degree $k=4$. Points show the mean $Q$ across $15$ graphs per $p$; shaded bands denote 95\% bootstrap confidence intervals over graph replicates.}
  \label{fig:modularity_vs_p}
\end{figure}

While it is reasonable to expect that injecting shortcuts should reduce community separation, whether $p_i$ yields a \emph{reliable, smooth, and usable} knob for modularity is not guaranteed a priori. Modularity is a global, partition-dependent statistic: the same set of shortcuts can disrupt community structure in nonlinear ways under finite size and different community-detection choices. We therefore treat smooth controllability as an empirical property of the generator and verify its existence directly. In our analyses, we operationalize modularity $Q(G_i)$ using the Newman--Girvan definition~\citep{newman2004finding} evaluated on a standard modularity-maximizing community partition~\citep{clauset2004finding}. Under this operationalization, $p_i$ behaves as a reliable, usable knob: modularity $Q_i$ nearly monotonically decreases as rewiring probability $p_i$ increases (Spearman $\rho=-0.92$, $p< 10^{-131}$; Kendall $\tau=-0.78$, $p< 10^{-89}$; Figure~\ref{fig:modularity_vs_p}). The decline is nonlinear (AIC$_{\text{quadratic}}=-171.422$ < AIC$_{\text{linear}}=-142.277$).

This yields an interpretable generative knob for semantic modularity: $p_i$ induces systematic variation in semantic modularity, which in turn is theoretically linked to creative performance through its influence on the reach of memory search.

\noindent\textbf{Robustness.} This relationship between $p$ and $Q$ is robust to walk length $T\in\{10, 20, 30, 50\}$, substrate size $|V|\in\{100,300\}$, degree $k\in\{4,10\}$, and 3 different seeds.

\subsection{Idea generation: a fixed semantic search process}
\label{sec:ideation}
Given a semantic network $G_i$, an agent generates ideas by performing stochastic semantic search from a prompt concept $s\in V$. We implement retrieval as a length-$T$ random walk on $G_i$, a parsimonious abstraction of spreading activation \citep{collins1975spreading,mednick1962associative}. Crucially, all agents share the same search process and hyperparameters. This keeps the model sharply identifiable: any systematic differences in ideation must arise from representational topology rather than agent-specific process variations.

A single walk produces a visited-node set $\mathcal{V}_i(s)\subseteq V$ and an induced trace subgraph $\mathcal{T}_i(s)$ over those visited nodes and edges. We treat $\mathcal{V}_i(s)$ as the agent’s idea trace in conceptual space: the number of distinct concepts visited,
$|\mathcal{V}_i(s)|$, is a minimal proxy for ideational breadth~\citep{hofstra2020diversity,uzzi2013atypical}, and the trace subgraph $\mathcal{T}_i(s)$ is the object exchanged in \textit{social interaction settings} explored later. 

\subsection{Exchangeable traces for social interaction}
\label{sec:compat}
A social creativity model requires that agents can respond to the same prompt and exchange what they generated in a commensurable form. Shared vocabulary ensures that a prompt $s$ denotes the same concept across agents. Shared substrate ensures that agents are comparable variants of the same semantic world. Together, these grounding choices mean that ideation traces are \textit{socially exchangeable} objects: $\mathcal{V}_i(s)$ and $\mathcal{T}_i(s)$ live in the same conceptual space for all agents and can therefore be compared (e.g., based on the extent of overlap), incorporated (e.g., augmentation of edges received through social `inspiration'), and analyzed at scale. 

\section{Individual Creativity Emerges from Semantic Network Structure}
\label{sec:emergence}
Before studying social effects, we verify the framework’s key precondition: creative behavior must emerge from constructed cognitive representations rather than being imposed as an agent attribute. Specifically, we test whether structural variation induced by the generative knob $p_i$---acting through semantic modularity $Q(G_i)$---produces systematic differences in ideational breadth under an identical retrieval process.

\subsection{Measure: ideational breadth under fixed retrieval}
\label{sec:emergence_setup}
We instantiate a population of semantic graphs $\{G_i\}$ as structured variants of a shared substrate $G_0$, as described above. Agents differ only in rewiring probability $p_i$ (restricted to the empirically monotonic modularity regime of the WS generator), which induces variation in modularity $Q(G_i)$. For each agent and prompt $s\in V$, we run $R$ independent length-$T$ random walks and record the number of distinct concepts visited,
\[
B_i(s)=|\mathcal{V}_i(s)|.
\]
We estimate each agent’s expected ideational breadth by averaging over $S$ sampled prompts and $R$ walk replicates,
\[
\widehat{B}_i=\frac{1}{S}\sum_{k=1}^{S}\left(\frac{1}{R}\sum_{r=1}^{R} B_i(s_k)\right),
\]
and quantify uncertainty by bootstrapping prompts $\{s_k\}_{k=1}^{S}$ (with replacement) and recomputing $\widehat{B}_i$.

\subsection{Result: modularity predicts exploratory reach}
\label{sec:emergence_result}

\begin{figure}[t]
  \centering
  \includegraphics[width=0.95\linewidth]{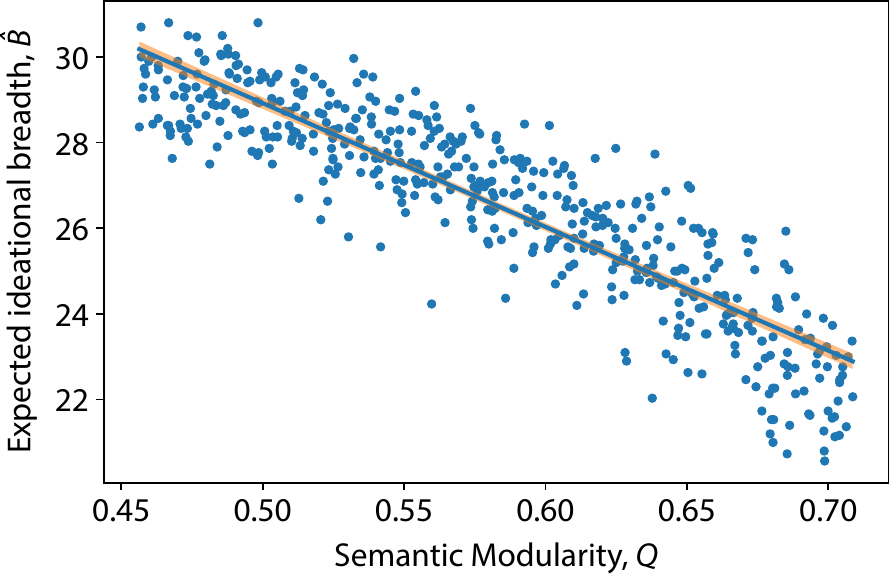}
  \caption{\textbf{Semantic modularity constrains exploratory access.} Each point is a semantic graph ($N_G=500$). Higher modularity $Q(G_i)$ predicts lower expected ideational breadth ($\widehat{B}_i$). We use $30$ walk replicates per sampled prompt per graph. Line shows the fitted linear trend; shaded band the 95\% CI.}
  \label{fig:B_vs_modularity}
\end{figure}

Across $N_G=500$ independently generated semantic graphs, modularity $Q(G_i)$ is strongly negatively associated with ideational breadth $\widehat{B}_i$ under the same walk-based retrieval procedure ($r=-0.90$, 95\% CI $[-0.91,-0.88]$, $p<10^{-179}$; Figure~\ref{fig:B_vs_modularity}). Rank-based estimates corroborate the monotonic relationship (Spearman $\rho=-0.91$, $p<10^{-192}$; Kendall $\tau=-0.73$, $p<10^{-131}$). A linear model indicates that higher modularity reduces breadth ($\widehat{B}\sim Q$: $\beta=-28.9$, 95\% CI $[-30.1,-27.7]$, $p<10^{-179}$, $R^2=0.81$). A robust Theil--Sen estimator shows that the effect is not driven by outliers ($\beta_{\mathrm{TS}}=-28.4$, 95\% bootstrap CI $[-29.9,-26.9]$). Model comparison suggests curvature beyond a linear trend (AIC$_{\text{linear}}=1441$ vs.\ AIC$_{\text{quadratic}}=1369.4$), showing diminishing returns as $Q$ approaches the low-modularity regime. 

These results show that the same single knob used to generate the population---rewiring probability $p_i$ acting through modularity---induces systematic individual differences in semantic exploration without changing how agents search. This pattern aligns with the intuition that highly modular networks trap traversal within local clusters, while cross-concept-community bridges facilitate access to concept regions that would otherwise be difficult to reach. Thus, individual creative potential emerges mechanistically from semantic topology, aligning with semantic-network accounts of creativity and providing an internal coherence check for the framework. 

\noindent\textbf{Robustness.} This relationship between $Q$ and $\widehat{B}$ is robust to walk length $T\in\{10, 20, 30, 50\}$, substrate size $|V|\in\{100,300\}$, degree $k\in\{4,10\}$, and 3 seeds.

\subsection{Implications for social modeling}
\label{sec:emergence_implication}
Importantly, the ability to control \textit{individual} creativity differences through a reliable knob $p_i$, combined with the \textit{shared} semantic grounding that provide socially exchangeable objects for ideation traces ($\mathcal{V}_i(s)$ and $\mathcal{T}_i(s)$; see the previous section), gives us the minimal conditions needed to study \textit{social creativity} phenomena without encoding them as assumptions. Because retrieval dynamics are fixed, subsequent dyadic and network-level effects cannot be attributed to tuned agent strategies or exogenous agent traits. Next, we proceed to examine whether theorized and empirically observed social creativity phenomena---cognitive stimulation from diverse peers and redundancy from shared sources---arise when agents exchange and incorporate ideation traces in the shared conceptual space.

\section{Cognitive Stimulation Emerges from Generative Social Interaction}
\label{sec:stimulation}
A central mechanism in social creativity theories is \emph{cognitive stimulation}: exposure to a partner’s ideas can open new associative routes, resulting in ideas that one could not generate on their own---especially when the partner borrows from conceptual neighborhoods that have low overlap with one's already-accessed regions~\citep{kelty2025innovation,baten2020creativity,paulus2007toward,paulus2000groups,dennis2003electronic,siangliulue2015toward,chan2016comparing}. We test whether cognitive stimulation emerges in our framework and whether its magnitude depends systematically on pre-interaction overlap.

\subsection{Dyadic interaction protocol}
\label{sec:stimulation_protocol}
We sample \emph{ordered} exposures $(i\!\to\! j)$ from the agent population constructed in the previous sections. For each ordered exposure, we sample multiple starting concepts $s\in V$ (prompts). For each $(i\!\to\! j, s)$, the source agent $i$ generates an initial ideation trace by running the fixed random-walk process on its semantic network, yielding a visited-node set $\mathcal{V}_i^{(1)}(s)$ and induced trace subgraph $\mathcal{T}_i^{(1)}(s)$. The recipient agent $j$ independently performs the same initial walk from the same $s$ on its own network, yielding $\mathcal{V}_j^{(1)}(s)$ and $\mathcal{T}_j^{(1)}(s)$.

We operationalize \emph{inspiration} as exposure to (and incorporation of) the source’s trace structure. Specifically, the recipient augments its edge set with associative links present in the source trace but absent from the recipient’s current network:
\[
E_j' \;=\; E_j \;\cup\; \bigl(E(\mathcal{T}_i^{(1)}(s)) \setminus E_j \bigr),
\]
yielding an updated network $G_j'=(V,E_j')$. The recipient then reruns the same random-walk ideation process from the same starting concept $s$ on $G_j'$, producing a second-round visited-node set $\mathcal{V}_j^{(2)}(s)$.

For each fixed $(i\!\to\! j, s)$ exposure, we run multiple independent walk iterations per round and summarize outcomes by averaging across iterations prior to statistical analysis. Each $(i\!\to\! j, s)$ thus contributes a single exposure-level mean, removing within-exposure stochasticity from the random walk process; remaining dependence across exposures sharing the same ordered
pair or prompt is accounted for in the regression analysis via fixed effects and clustered standard errors.

\subsection{Measures}
\label{sec:stimulation_measures}
We quantify (i) \emph{pre-interaction overlap} between the source and recipient’s initial ideation, and (ii) the recipient’s \emph{stimulation benefit} after incorporating the source trace. Pre-interaction overlap is measured as Jaccard similarity in round one:
\[
\mathrm{Overlap}_{i\to j}(s) \;=\;
\frac{|\mathcal{V}_i^{(1)}(s)\cap \mathcal{V}_j^{(1)}(s)|}{|\mathcal{V}_i^{(1)}(s)\cup \mathcal{V}_j^{(1)}(s)|}.
\]
Stimulation benefit is measured as the number of \emph{new} concepts the recipient accesses in round two relative to its round-one:
\[
\Delta_{i\to j}(s) \;=\; \bigl|\mathcal{V}_j^{(2)}(s)\setminus \mathcal{V}_j^{(1)}(s)\bigr|.
\]
For each exposure $(i\!\to\! j, s)$, we compute $\mathrm{Overlap}_{i\to j}(s)$ and $\Delta_{i\to j}(s)$ for each walk iteration and then average over iterations, yielding per-exposure means used in statistical tests.

\subsection{Result: overlap predicts marginal benefit}
\label{sec:stimulation_result}

\begin{figure}[t]
  \centering
  \includegraphics[width=0.75\linewidth]{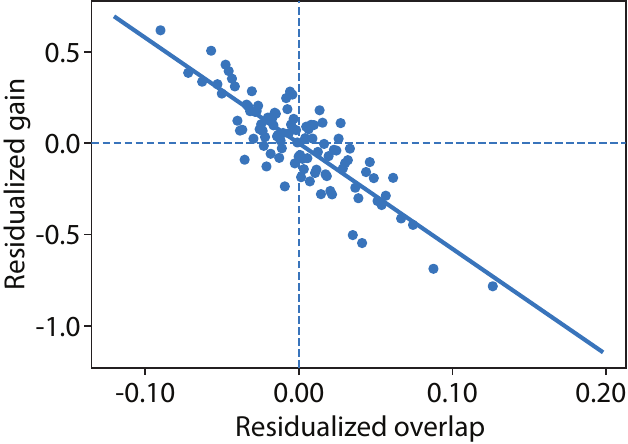}
  \caption{\textbf{Lower pre-interaction overlap predicts larger stimulation gains.} We plot the fixed-effects partial relationship between pre-interaction overlap and stimulation benefit after residualizing both variables by ordered-pair fixed effects and prompt fixed effects. We partition residualized overlap into $100$ quantile bins and plot, for each bin, the mean residualized overlap against the mean residualized gain. The fitted line has a slope equal to the estimated $\beta$ from Eq.~\ref{eq:stimulation_fe}.
    }
  \label{fig:overlap_vs_gain}
\end{figure}

Because each ordered pair $(i\!\to\! j)$ reappears across many prompts, exposure-level observations are not independent: the same pair shares stable structural properties that could jointly shape both overlap and gain. To isolate the within-pair relationship, we estimate a two-way fixed-effects model with ordered-pair and prompt fixed effects:
\begin{equation}
\label{eq:stimulation_fe}
\Delta_{i\to j}(s)
\;=\;
\beta\,\mathrm{Overlap}_{i\to j}(s)
\;+\;
\alpha_{i\to j}
\;+\;
\gamma_{s}
\;+\;
\varepsilon_{i\to j,s},
\end{equation}
where $\alpha_{i\to j}$ absorbs all pair-specific time-invariant factors (including both agents’ fixed semantic structures and directionality), and $\gamma_s$ absorbs prompt-specific difficulty or baseline exploration differences shared across pairs. We report standard errors clustered by ordered pair to account for residual dependence across prompts within the same $(i\!\to\! j)$.

We observe a clear negative relationship between pre-interaction overlap and stimulation benefit: $\hat{\beta}=-5.8$ (95\% CI $[-6.7,-4.9]$, $p< 10^{-34}$; $4{,}998$ exposures across $500$ ordered pairs; $R^2=0.50$; Figure~\ref{fig:overlap_vs_gain}). Interpreted on the overlap scale, a $0.10$ increase in Jaccard overlap corresponds to $0.58$ fewer newly accessed concepts on average, holding the ordered pair and prompt fixed. Thus, exposures in which the source and recipient initially traverse less-overlapping conceptual regions yield larger gains in newly accessed concepts after inspiration, whereas higher-overlap exposures yield smaller gains.

This effect is mechanistic, as it is not produced by any reward term, diversity bonus, or adaptive strategy. Agents run the same fixed search process before and after inspiration, and the update rule only adds novel associative links from a partner’s trace. Cognitive stimulation arises because low-overlap partners contribute bridges into genuinely new semantic regions, whereas high-overlap partners contribute fewer structurally novel links capable of redirecting subsequent exploration.

\noindent\textbf{Robustness.}
The negative overlap--benefit relation persists under variations in walk length $T\in\{{10, 20, 30}\}$ and $3$ seeds.

\section{Social Network-Level Redundancy Emerges from Shared Sources of Inspiration}
\label{sec:redundancy}
Cognitive stimulation is often framed as a dyadic benefit, but collective creativity unfolds in social networks where multiple recipients may draw inspiration from the same source. This creates a structural tension: a highly creative source can help many individuals expand their search, yet the same shared source can also induce convergence among recipients—yielding \emph{redundancy} as a network-level externality~\citep{kelty2025innovation,baten2020creativity}. Rather than encoding this tension through explicit costs, we test whether redundancy arises under a minimal model: when recipients incorporate traces in a shared conceptual space, does \emph{sharing an inspiration source} increase recipient--recipient overlap even when recipients do not interact or adapt to each other?

\subsection{Controlled design: shared vs.\ independent sources}
\label{sec:redundancy_design}
We compare two conditions that differ only in whether two recipients share an inspiration source. We sample two \emph{sources} to be highly creative under our generative construction (i.e., with bottom $20$\% modularity) and two \emph{recipients} to be low-creative (i.e., top $20$\% modularity). All agents ideate on the same prompt $s\in V$ using the same fixed random-walk process. Recipients then incorporate edges derived from the source trace(s) and ideate again, yielding post-inspiration traces.

\noindent\textbf{Independent-source }(\textsc{Control}).
Recipient $a$ incorporates edges derived from source $h_1$’s trace, while recipient $b$ incorporates edges derived from source $h_2$’s trace.

\noindent\textbf{Shared-source} (\textsc{Triad}).
Both recipients incorporate edges derived from the same source $h_1$’s trace, holding all other aspects constant.

This construction yields matched \textsc{Triad}--\textsc{Control} comparisons for each $(h_1,a,b)$ instance, differing only in whether the inspiration source is shared.

\subsection{Measures}
\label{sec:redundancy_measures}
We quantify redundancy as similarity between the two recipients’ \emph{post-inspiration} visited-node sets. Let $\mathcal{V}_a^{(2)}(s)$ and $\mathcal{V}_b^{(2)}(s)$ denote recipients’ second-round visited-node sets for prompt $s$. We measure recipient--recipient overlap as:
\begin{equation*}
\label{eq:redundancy_jaccard}
R(s)
\;=\;
\frac{|\mathcal{V}_a^{(2)}(s)\cap \mathcal{V}_b^{(2)}(s)|}{|\mathcal{V}_a^{(2)}(s)\cup \mathcal{V}_b^{(2)}(s)|}.
\end{equation*}
For each matched instance, we define the paired redundancy difference as
$\Delta(s)=R_{\textsc{Triad}}(s)-R_{\textsc{Control}}(s)$.

\subsection{Result: shared inspiration increases redundancy}
\label{sec:redundancy_result}

\begin{figure}[t]
  \centering
  \includegraphics[width=0.9\linewidth]{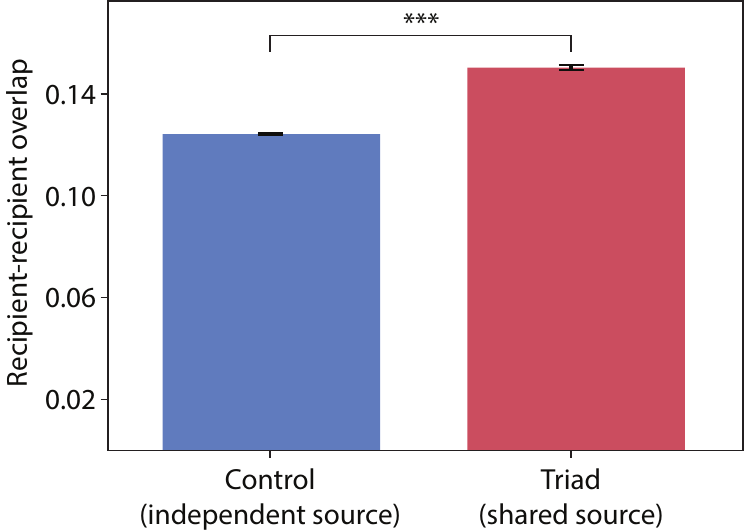}
  \caption{\textbf{Shared inspiration increases redundancy in recipients’ semantic exploration.} Bars show mean post-inspiration recipient--recipient overlap across $495{,}000$ matched instances; whiskers denote 95\% CI computed with SEs clustered by inspiration source; ***$p<0.001$.
}
  \label{fig:shared_source_redundancy}
\end{figure}

Because many instances share the same inspiration source, comparisons are not independent within source; we therefore base inference on a source-clustered estimator:
\begin{equation*}
\label{eq:redundancy_cluster}
\Delta_{h} \;=\; \alpha \;+\; \varepsilon_{h},
\end{equation*}
where each observation is a matched instance and standard errors are clustered by source $h$. We find that across $495{,}000$ matched instances, shared inspiration sources in \textsc{Triad} significantly increases redundancy in recipients' semantic exploration than \textsc{Control} ($t=55.9$, $p<10^{-300}$; Figure~\ref{fig:shared_source_redundancy}). The paired mean difference is $\Delta=0.026$ (95\% CI $[0.025,\,0.027]$), corresponding to a large effect size (Cohen’s $d_z=1.4$).

Mechanistically, this redundancy is not imposed by any explicit penalties. Recipients do not adapt to one another. The increased overlap arises solely from shared structural updates: a common source introduces similar new bridges into both recipients' semantic networks, biasing subsequent walks toward overlapping regions.

\noindent\textbf{Robustness.}
The redundancy effect persists under variations in walk length $T\in\{{10, 20, 30}\}$ and $3$ different seeds.

\section{Discussion}
\label{sec:discussion}

This paper’s core contribution is to make collective creativity formally tractable under a \emph{generative} view of cognition, instead of modeling creativity as an exogenous agent trait assigned at the social level. We show that a single semantic structural knob can induce creative ability differences across agents. This way, canonical social-creativity effects become well-posed and arise organically. Long-standing claims in social-creativity theory then become mechanistically testable questions: when does exposure expand exploration, when does it induce convergence, and how do these effects depend on the structures of minds and of networks?

The framework primarily opens a counterfactual and hypothesis-generation space. Because agent differences are induced by a controlled structural dimension of semantic memory, one can systematically vary cognitive mechanisms that shape how far individuals are willing or able to explore. Demographic and developmental shifts (e.g., age-related changes~\citep{ramscar2017mismeasurement,wulff2022structural,campidelli2026creativity}), as well as constructs such as absorptive capacity, cognitive control, and motivation~\citep{miguelez2015knowledge}, can be instantiated as parameters that gate traversal and incorporation \emph{without} abandoning representational grounding. Our present model is intentionally minimal, as all agents run the same search process and inspiration is implemented as full incorporation of a partner's trace. But precisely because the baseline is clean, extensions become interpretable: partial uptake (absorptive capacity), selective attention to specific trace regions, decay and interference over time, or goal-biased traversal can be introduced to test when stimulation saturates, reverses, or becomes fragile.

At the social level, the same cognitive foundation enables principled questions about how interaction structure shapes creative outcomes. Tie formation dynamics (e.g., homophily on demographic or ideation success cues) can be encoded to study how teams self-assemble, how network evolution alters the stimulation--redundancy tradeoff, and when ``spreading out'' inspiration sources across the network outperforms repeatedly drawing from a small set of highly generative hubs~\citep{page2019diversity,baten2024ai,bangash2025musescorer}. Likewise, team size and temporal exposure schedules can be varied to ask when diversity is beneficial versus when shared sources accelerate convergence~\citep{wu2019large,kamal2025beginner}. While we did not model strategic partner choice, incentives, or endogenous rewiring here, these are natural next layers: the framework provides a mechanistic sandbox in which such social processes can be tested against cognitive constraints rather than assuming them as abstract payoffs.

\section{Code and Data Availability}
The analysis code and data can be found here: \\https://github.com/cssai-research/gen-collective-creativity.
\printbibliography
\end{document}